\documentstyle{caosp}
\begin{document}
\htitle{IR variability of CP stars}
\hauthor{F.A. Catalano {\it et al.}}
\title{On the near infrared variability of chemically peculiar stars
\thanks{Based on observations collected at the European Southern 
Observatory, La Silla Chile.} } 
\author{F.A. Catalano \inst{1,} \inst{3} \and F. Leone \inst{2,} \inst{3}}
\institute{Istituto di Astronomia dell'Universit\`a di Catania \and 
Osservatorio Astrofisico di Catania \and C.N.R.-G.N.A. Unit\`a di 
ricerca di Catania, Italy}
\maketitle
\begin{abstract}
Some CP stars have recently been discovered by Catalano et al. (1991) to 
be variable also in the near infrared, although with smaller amplitudes 
than in the visible. Hence an observational campaign was started in 
which the infrared light variability of a number of CP2 and CP4 stars 
has been investigated at the ESO-La Silla Observatory in the bands $J$, 
$H$, and $K$. As a general result, infrared variations show the same 
behavior in all three filters but amplitudes are smaller than in the 
visible. 
\end{abstract}

\section{Introduction}
\label{intr}
Kroll et al. (1987) showed that the near infrared fluxes and colors of 
Chemically Peculiar stars (or CP stars, according to Preston's (1974) 
scheme), when compared to a black body, are normal, like that of early 
main sequence stars. IRAS data could even prove that the normality of IR 
fluxes is guaranteed to at least 25$\mu$ (Kroll 1987): only two CP4 stars 
showed flux excesses longward of 60$\mu$, showing cold circumstellar 
material, which is not uncommon among early B stars. Moreover Leone \& 
Catalano (1991) have shown that the solar composition Kurucz model 
atmospheres, which are used to fit the spectra of CP stars from $\lambda$5500 
to $\lambda$16500~\AA, give a fair representation of the overall flux 
distribution, with the exception of the Balmer region, where CP stars 
appear generally brighter than normal, this excess being just a few 
percent of the total flux. \\ 
\begin{table}[t]
\footnotesize
\begin{center}
\caption{The CP stars checked for variability in the near infrared.}
\label{t1}
\begin{tabular}{rrrrrr}  \hline\hline
 SrCrEu & HD~~~3980 & HD~~24712 & HD~~49976 & HD~~72968 & HD~~83368 \\
      & HD~~96616 & HD~~98088 & HD~101065 & HD~111133 & HD~118022 \\
      & HD~125248 & HD~126515 & HD~137949 & HD~148898 & HD~153882 \\
      & HD~164258 & HD~203006 & HD~206088 & HD~220825 & HD~221760 \\
Si et al. & HD~~10783 & HD~~12447 & HD~~74521 & HD~~90044 & HD~116458 \\ 
      & HD~119419 & HD~125630 & HD~147010 & HD~166469 & HD~170397 \\
      & HD~187473 & HD~223640 &  &  &  \\
Si & HD~~12767 & HD~~19832 & HD~~25267 & HD~~29305 & HD~~37808 \\
   & HD~~54118 & HD~~56455 & HD~~66255 & HD~~73340 & HD~~92664 \\ 
   & HD~114365 & HD~116890 & HD~122532 & HD~124224 & HD~133880 \\
   & HD~144231 & HD~145102 & HD~203585 & HD~221006 &  \\
He weak & HD~~~5737 & HD~~22470 & HD~~28843 & HD~~35456 & HD~~37151  \\
  & HD~~49333 & HD~~74196 & HD~125823 & HD~137509 & HD~142990 \\
  & HD~144334 & HD~148199 & HD~168733 & HD~175362 &    \\
He rich & HD~~36485 & HD~~37017 & HD~~37479 & HD~~37776 & HD~~59260 \\
   & HD~~60344 & HD~~64740 &  &  &   \\  \hline\hline
\end{tabular} \end{center} \end{table}
\vspace{-2mm}

However, in spite of this normal infrared behavior, peculiar abundances 
and/or magnetic fields seem to affect the near infrared too; in fact, 
Catalano et al. (1991) have shown that, out of the eight CP stars 
monitored throughout their rotational periods, at least six are variable 
in the near infrared, although with smaller amplitudes than in the 
visible. This unexpected result led us to start an observational 
campaign aimed at searching for infrared variability and also to better 
understand the origin of the light variability, which is one of the 
outstanding observational aspects of these stars. 

\vspace{-3mm}
\section{Observations}
\label{obse}
\vspace{-2mm}
The observations have been carried out in the near IR bands J, H, and K 
at the 1-m photometric telescope at ESO, La Silla, Chile, using an InSb 
detector cooled with liquid nitrogen. A detailed description of the ESO 
infrared photometers can be found in Bouchet (1989).

The data have been collected during several observing runs from July 1986 
through January 1993. All program stars were measured relative to closeby 
comparisons, which were chosen to have as similar color and brightness as 
possible. The integration times, the number of cycles, and the desired 
r.m.s. accuracy in the mean level were optimized to get a 2\% maximum 
error in the observations: the resulting accuracy in the final reduced 
data is typically 0.006 mag. ESO standard software was used for all 
reduction steps. Magnitudes in the standard IR system have also been 
obtained by observing suitable standard stars from the ESO list (Bouchet 
et al. 1991).

The adopted ephemeris elements of the infrared light curves for the 
programme stars have been mainly taken from Catalano \& Renson (1984, 
1988, 1997), Catalano, Renson \& Leone (1991, 1993), and references 
therein. The results concerning the SrCrEu and Si et al. stars have 
been published elsewere (Catalano et al. 1997, 1998). 

\vspace{-2mm}
\section{Discussion and conclusions}
\vspace{-2mm}
Near infrared variability has been found to be present in the large 
majority of the CP2 stars studied. The typical trend of CP2 stars to 
present smaller amplitude light variations at increasing wavelength 
is confirmed: the amplitudes in the near infrared are smaller than 
in the visible. In most cases the variations have been found to show 
very similar behavior and in phase with each other in all filters.

In a previous paper (Catalano et al. 1991) we investigated the effects 
of high metallicity at the near infrared wavelengths and showed that a 
Kurucz model atmosphere with a metal content ten times the solar one 
could explain a three percent variation in the near infrared brightness, 
which is the typically observed value.

The influence of the magnetic field in the atmosphere structure has 
been quantitatively discussed by some authors in some particular 
configurations, however the most general approach has been carried out 
by Stepien (1978) who showed that, according to the direction of the
toroidal electric currents in the outermost layers, the star's shape can be 
prolate or oblate with respect to the magnetic axis: the differences 
between the polar and equatorial values of the radius being up to 3\%. 
The results obtained by Stepien lend support to a distorted figure of the 
star up to a few percent and to small variations (2-3\%) of the effective 
temperature over the surface, which in some cases, can contribute to 
the observed light variations. While this explanation is not valid as 
far as it concerns the visible light variations of many CP stars, 
because of the different behaviours presented by the $u$, $v$, $b$, 
and $y$ curves, it cannot be excluded that the non-spherical shape 
of the star as seen at the infrared wavelengths could contribute to the 
observed variability, since the magnetic pressure importance increases 
in the outer layers.

After completing the analysis of our infrared data, we hope to be able to 
disentangle the relative contributions of these two mechanisms from 
the study of the phase relation between the magnetic field and infrared 
variations.

\end{document}